\documentclass[aps,prl,preprint,showpacs, groupedaddress]{revtex4}
\usepackage{graphicx}

\begin{document}

\title{Direct Observation of a Cross-Field Current-Carrying Plasma
\\ 
Rotating Around an Unstable Magnetized Plasma Column
 }
\author{S. Jaeger, and Th. Pierre}

\affiliation{Laboratoire PIIM, CNRS and
Aix-Marseille Universit\'e, Campus St Jerome, F-13397 Marseille \normalfont\large
Cedex 20, France}

\begin{abstract}
The low-frequency instability of a magnetized plasma column and 
the anomalous radial plasma convection are shown to be linked with 
the existence of a radial cross-field current along a plasma channel
 rotating around the central magnetized plasma. 
 The direct observation of the rotating plasma is obtained
using an ultra-fast intensified camera.
The ionizing electrons injected along the axis of the plasma column 
contribute to the accumulation of negative charges when the axial 
collector is at floating potential. 
The required neutrality leads to the continuous radial expulsion of 
energetic electrons and to the formation of a rotating plasma channel (m=1 
unstable wave). 
\end{abstract}


 \maketitle

   The problem of anomalous transport in magnetized plasmas has
been investigated during the last twenty years. Historically, 
the evidence that plasma diffusion at the edge of a 
magnetized plasma column is anomalous dates from the early 
sixties. The russian research in confinement of hot plasmas \cite{ioffe01} 
reported first the fact that plasma filaments escape from the central 
magnetized plasma and reach the wall, leading to the degradation of the quality 
of the confinement. This has been later mentioned by Kadomtsev
 \cite{kadom01} and also by Chen \cite{chen67} with emphasis on the problem
of the leakage of fusion devices.
During the last fifty years, many investigations have been devoted to 
the identification of the various mechanisms leading to the 
destabilization of a magnetized plasma column. In fact, the most 
efficient mechanisms involve the creation of an azimuthal separation of
charges at the edge of a magnetized plasma \cite{simon01}. Various phenomena can lead
to this situation such as centrifugal effects, the
curvature of the field lines, or the collisional slow electric drift of the ions. 
Moreover, finite ion Larmor radius effects have to be taken into account when 
the magnetic field is low.

We have investigated the cross-field transport of the plasma
in the edge of a magnetized plasma column and we have obtained
the experimental evidence that the plasma detected around the plasma
column is not convected directly from the central plasma.
We have observed for the first time a rotating plasma around a 
magnetized plasma column using an ultra-fast intensified
camera and we conclude that the plasma is locally created by the ionizing 
electrons drifting outside the central column through electric drift.
We show also that the existence of a radial electronic 
current across the rotating plasma channel is the key parameter for the destabilization 
of the plasma column. In fact, when the impedance of the axial collector is high,
the non-exact balance between axially injected electrons and 
axially collected electrons determines in return a radial current that is 
made possible through radial electric drift inside the large azimuthal electric field
of an unstable wave.

To the best of our knowledge, this situation involving a
 non-exact balance between the injected flux of negative charges and 
 the collected axial current has not been yet carefully investigated.
 Our experimental situation is very similar to the experiments by
Chen and Cooper on instabilities in a reflex discharge 
\cite{chen62a, chen62b} where the turbulent radial current towards the cylindrical anode 
was necessary to operate the discharge. 
More recently, Rypdal et al. \cite{rypdal97} have shown 
that ion polarization current was necessary to establish a 
fluctuating equilibrium in a toroidal magnetized plasma. 
Cross-field currents are also often involved in the description of 
intermittency and transient events in space plasmas, in solar physics 
\cite{space, solar}, and also in the case of plasma 
thrusters \cite{thruster}. Finally, due to charge injection, our experiment has similarities
 with the experiments in non-neutral plasmas exhibiting the diocotron instability \cite{delzanno}. 
Our findings could lead to a different analysis of the
anomalous transport in collisional laboratory and fusion plasmas.

The results reported here are obtained in the magnetized
plasma produced in the MISTRAL device \cite{art12, art13}. A 
thermionic discharge with 32 hot tungsten filaments is operated 
in a large source chamber connected to a cylindrical vessel (40 cm in diameter, 
1 meter in length). 
The vessel is inserted inside a solenoid made of 20 water-cooled 
coils equally spaced along the cylindrical vacuum chamber. The 
maximum magnetic field strength is 0.03 T.
The potential of the anode inside the source plasma determines
the plasma potential inside this plasma. This potential is 
the reference value for ionizing electrons produced inside the source chamber and 
entering the plasma column. 
A very fine mesh grid at the entrance section of the
column allows an electric insulation between the source plasma 
and the target plasma. The grid is kept at floating potential. 
Only energetic electrons overcome the potential of the grid 
(typically -15 volts) and enter the column through a 8 cm circular aperture. 
A linear magnetized plasma column is produced (8 cm diameter, 90 cm length)
inside the cylinder.
The vacuum chamber is evacuated at a pressure of about $10^{-6}$ mbar.
The collector at the end of the plasma column is made of a copper plate
that can be replaced by a fine metallic mesh grid allowing optical
measurements along the axis of the device.
The plasma column is surrounded by an insulated collecting cylinder 
-20 cm in diameter- divided into two separate half-cylindrical 
parts. Each half-cylinder can be biased independently and the temporal 
evolution of the collected current is recorded. The evolution of the 
transverse current is measured when several parameters are changed i.e. 
the potential of the axial collector that is varied from floating to collecting 
polarization, the working gas pressure and the potential of the anode.
The radial density profile is almost gaussian with
a central density in the range $10^{15}$ to $10^{16} m^{-3}$. 
The electron temperature ranges from 2 to 4 eV depending on the 
working gas pressure \cite{art13}, ranging from $10^{-5}$ to $10^{-3}$ mbar 
in argon.
Several Langmuir probes consisting in 2 mm tips protruding
outside a semi-rigid coaxial cable are used for radial scan of the
plasma density, the plasma potential and the electron temperature. 
The electron density fluctuations are recorded by three digital 
scopes and the power spectrum of the fluctuations is monitored using 
a spectrum analyzer.

The radiation from the plasma is used as a fast and efficient
diagnostic \cite{zweben}. The camera is a fast gated and intensified model.
The intensifier incorporates a S25-type photocathode that is 
sensitive to the bright near-infrared line at 810 nm
of the excited argon atoms. The axial probing by Langmuir probes has shown that the
density fluctuations are uniform along the axis of the plasma column. 
This flute character of the instability allows the axial imaging of the whole plasma. 
The central part of the magnetized plasma column is very bright and a disk is 
inserted to mask this region of the plasma.
When a coherent unstable wave is established,
a trigger signal is built by using the density fluctuations recorded by
a Langmuir probe located at the edge of the central plasma column.
A variable delay is added before triggering the camera in order to 
produce a stroboscopic analysis of the rotating plasma.

The second specific diagnostic used in our measurements is based on
a radially movable electrostatic energy analyzer (Retarding-Field Analyzer, RFA)
that can be oriented in order to face
the source plasma or alternately to face the end-plate. Measurements 
perpendicular to the B-field are not possible due to the strong magnetization
of the electrons.  The electron parallel velocity distribution function 
is obtained using this RFA that is 
made of a small collector (14 mm in diameter) behind two high transparency grids.
The first grid is kept at a potential slightly higher than the plasma
potential. The second grid is the selecting grid. It is swept in 
polarization from -50 volts to +10 volts. The collector of the RFA is grounded through 
a resistor allowing to measure the collected current and hence the
energy distribution of the electrons along the direction parallel to 
the magnetic field. After rotation of one half-turn, the direction of 
anaysis can be reversed from
the parallel upstream direction to the parallel downstream. This gives
very valuable information on the asymmetry of the distribution function.

The column is most often unstable when the plasma
discharge is switched on. The magnetic field is 0.017 T corresponding
to a cyclotron frequency of 6.46 kHz for argon ions. Previous studies have shown that the 
polarization of the collector is
an important parameter for the destabilization of the plasma column
\cite{klinger97, gravier99} though a definitive explanation of
the destabilization has not yet been obtained. The pressure is fixed 
at  $2.10^{-4}$ mbar. 
The frequency of the instable wave is about 4 kHz. At a fixed gas pressure, 
two crucial parameters are identified : the biasing of the source plasma and 
the impedance of the end-plate. A larger flux of ionizing electrons
is injected when the potential of the source anode is lower. On the other hand,
a low impedance of the end-plate connected to the ground across a variable resistor
would lead to a quasi complete collection of the primaries. 

On the contrary, a high impedance of the collector leads to
the build-up of a negative sheath and to the floating potential of the collector.
We note that changing the load of the collector from short-circuit (stable plasma)
to high impedance (unstable plasma) fully controls the onset of the instability.
This particular point leads to the possibility to make a fruitful analogy between
our experiment and experiments in non-neutral plasmas \cite{davidson}.
Indeed, inside a Penning trap, the diocotron instability is recorded due to the
reflexion of the electrons at each end of the trap and to the existence of the
modulated self-consistent radial electric field.

We have chosen to set the experiment into the most unstable regime 
which is obtained with a floating collector. The potential
of the anode in the source plasma is Va=-5V.  The time-series of the density
 fluctuations recorded by the Langmuir probe at the edge of the 
central plasma column is shown in Fig. 1(a). 
The shape of the signal ressembling to a cnoidal wave indicates the strongly
nonlinear state. This is confirmed by spectral analysis.
The study of the correlation with
an other probe located at a different angular position exhibits clearly
the m=1 structure of the mode. The relative 
fluctuation level is maximum at the edge of the plasma column. The frequency
of the unstable mode is always below the ion cyclotron frequency and
increases with decreasing the potential of the anode of the source chamber,
i.e. increasing the flux of the ionizing electrons entering the cylindrical
magnetized target plasma.

The optical probing using the fast camera gives a very valuable insight
on the stucture of the instability. We have measured that the most 
intense radiation is produced by near-infrared atomic lines.
Successive frames are recorded with an
increasing delay after local detection by the probe of the maximum of the 
plasma density.
A video file is obtained exhibiting the rotation of a
bright plasma channel around the central magnetized plasma
column. The plasma channel has a large extension along the plasma
column and establishes a connection between the central plasma and the wall. 
A series of successive frames is depicted in Fig. 2 where the bright central part of 
the column is masked. If the collector is grounded, 
the rotating plasma channel
suddenly disappears and the central plasma is stable.

The measurement of the temporal evolution of the current collected 
by the two half-cylinders at the wall is also highly instructive : Fig. 1(b) shows the
current (electron current) collected by one of the half cylinders. The
square shape of the signal corresponds to the successive collection
by the half-cylinders. It is in complete agreement with the frames
depicted in Fig. 1(a). 
The time-series of the second cylinder is opposite
in phase and is therefore not displayed. The most important conclusion is that
a negative current is continously drawn radially from the central
magnetized plasma column. Obviously, this current is part
of the axial current across the first section of the plasma column.
The required global neutrality implies a radial flux of negative charges.
However, it is important to note that the mechanism for this cross-field 
current is not so obvious. It is highly probable that the azimuthal electric field
across the m=1 structure leads to the radial ExB drift of the electrons present in
the edge of the plasma column, including the ionizing electrons.
In return, this ExB convection builds a plasma channel because ionizing electrons
 are convected. Finally, the plasma channel connects electrically 
 the core plasma to the wall.
This localized plasma sheet is ExB rotating around 
the central column under the global influence of the radial electric field. 

The radial current flowing transversally is measured. It is
varied by changing the flux of the axially injected ionizing 
electrons or by collecting or not the ionizing electrons at the end of the 
plasma column. Thecrude estimation of the collisionnal transverse current is 
far below the measured value of the current density across the magnetic field.

The experiment shows that the collisonality is an important factor but 
more measurements have to be conducted in
order to get a better understanding. For instance, we measure that the
plasma potential inside the central column increases from -8 volts to -2 volts
when the potential of the collector is continously changed from floating
potential to ground. The frequency of the unstable mode decreases
and a turbulent state is obtained. To summarize, the flux of
ionizing electrons with incomplete collection on the axis leads to a coherent
unstable wave with transverse current. Decreasing the flux of injected negative charges
or increasing the parallel collection leads to the  decrease in frequency and to a
lower radial current.

\begin{figure}
\includegraphics{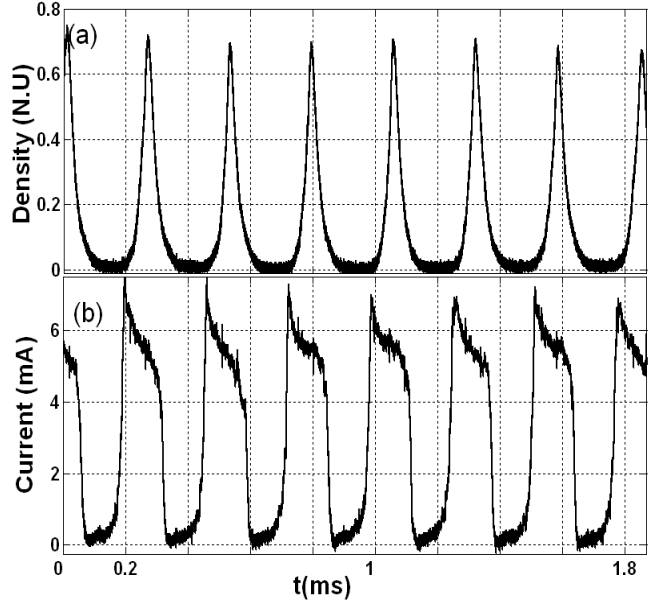}
\caption{\label{fig:Spectre} Temporal evolution of the density and 
radial current: (a) Plasma density 
recorded by a Langmuir probe located behind the limiter, normalized to 
the density on the axis of the plasma column; (b) Electronic current
recorded on one of the collecting half-cylinders. The current is present when the 
plasma channel rotates in front of the corresponding cylinder.
}
\end{figure}

\begin{figure}
\includegraphics{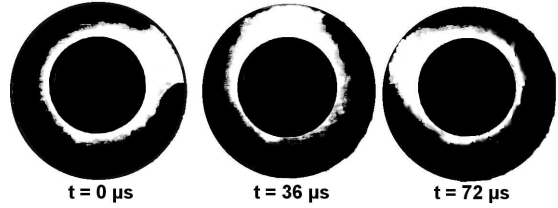}
\caption{\label{fig:RotationBras} Axial view of the plasma column whose bright
central part has been masked. The successive frames recorded by the 
ultra-fast intensified camera exhibit the rotation of the plasma channel 
establishing the electrical connection between the central plasma and the collecting cylinder. The rotation is counter-clockwise and it
is consistent with the measured inward radial electric field.}
\end{figure}

\begin{figure}
\includegraphics{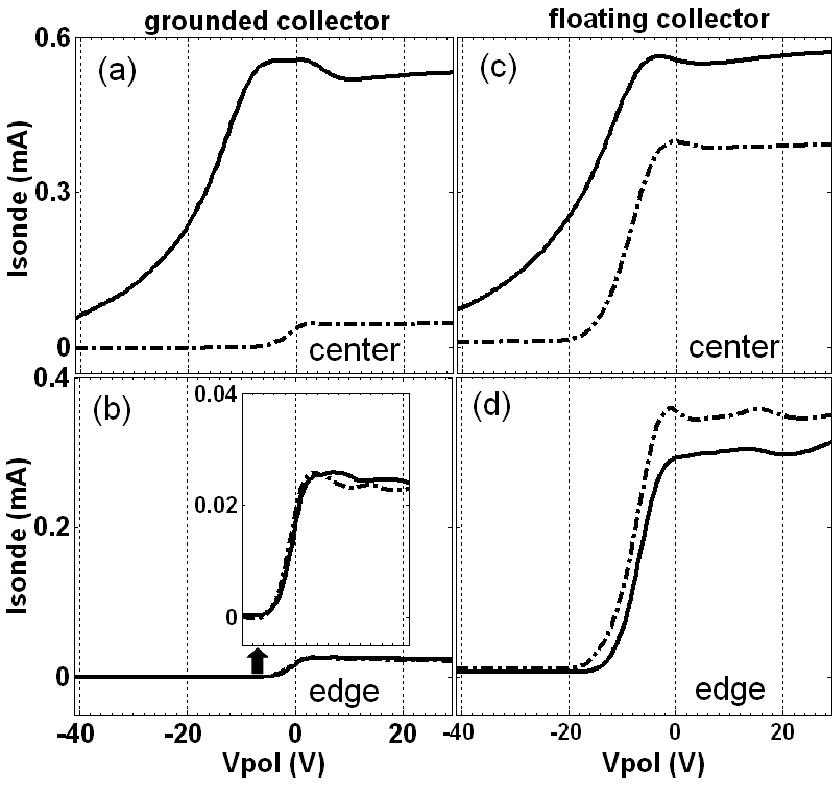}
\caption{\label{fig:CaracSG} Integrated energy distribution of the electrons 
obtained from the energy analyzer with grounded (a, b) or floating collector (c,d), in the central plasma (a,c)
and behind the limiter near the wall (b,d). 
Solid line: RFA oriented toward the source. Dash-dot 
line: RFA oriented toward the collector. It is clear that energetic electrons are detected outside the core
plasma only when the collector is floating. We note that a large number of electrons flowing upstream
is also observed in the case of floating collector.}
\end{figure}

At first glance, one could estimate that no ionization process exist in the shadow of the limiter. 
This is not the case as it is seen by measuring the velocity distribution of the electrons is measured 
in that region. The presence of ionizing electrons around the core plasma, i.e. in the 
shadow of the limiter (SOL), is investigated using the RFA in various conditions
and at several radial locations from the wall 
to the edge of the central plasma column. In the SOL the data acquisition is performed only
when the rotating plasma channel is at the location of the analyzer.
Moreover, the Electron Energy Distribution Function (EEDF)
is measured upstream and downstream. In the latter case, electrons reflected by the floating
axial collector can be detected.The results are summarized in Fig. 3. 
At first, when the axial collector is grounded, the measurement of the 
EEDF is performed on the axis of the central
magnetized plasma column. With the RFA facing the source plasma (upstream) 
the solid line in Fig. 3(a) exhibits a large distribution in energy of the collected electrons
(the EEDF would be obtained from the first derivative of the collected current for potentials lower
than the potential corresponding to the saturation current) 
and a large ionizing tail beyond 20 eV. With the RFA facing to the collecting
end-plate, a cold low-density distribution is recorded. The central plasma column 
is stable and quiet. When the RFA is located near the wall, a very cold and tenous plasma
is detected [Fig.3(b)].

On the other hand, with a floating collector, the situation is quite different. 
The integrated EEDF is displayed in Fig. 3(c) with the RFA located in the center of the plasma
column. When the RFA collects electrons coming directly from the source plasma (solid line),
the integrated EEDF is somewhat similar to the case with grounded collector, but when the RFA
is oriented toward the collector, an energetic tail is still recorded (dashed line).
In fact, the end-plate at floating potential reflects a large part of the electron
distribution as seen in Fig. 3(c).
The most important point is that changing the radial location of the RFA to the
edge of the cylinder (2 cm from the wall), the measurement of the EEDF shows in Fig. 3(d) that
a large number of energetic electrons are still present and that the distribution
is similar upstream and downstream. This explains the existence of the plasma channel
across the magnetic field. To the best of our knowledge, this observation is quite new.

In the case of a laboratory plasma with a low ionization degree
and rather high collisionality, the plasma detected transiently outside the column
is due to the local ionization by drifting energetic electrons. 
In evaluating the final transformation of the parallel current 
into transverse current, the detailed mechanims of
the axial collection of ions and electrons have to be taken into account, as well as the
recombination processes inside the plasma.
This detailed experimental study is important for a better understanding of the transport 
of plasma across the magnetic field.

In summary, we have presented an experimental study of a low-frequency instability
in a magnetized plasma column leading to the evidence that it is 
due to the necessity
to evacuate radially the axially injected negative charges following a mechanism very
similar to the diocotron instability.
In fact, the electrons are ejected due to the 
build-up of a large amplitude m=1 deformation of the column. The 
ejected energetic electrons
create a transverse plasma channel around the core plasma. 
In parallel, the radial electric field is still effective inside the
plasma channel and it leads to the azimuthal ExB drift 
of the electrons and hence to the rotation of the structure. 
In conclusion, the key parameter is the radial electric field, but the
effective parameters are the pressure of the gas (collisionality) and the
potential at the collector determined by an external polarisation
or alternately by its impedance. More precisely,
the impedance of the axial collector defines the
balance between injected and axially collected charges. This
determines in return the radial electric field and the radial
current when the nonlinear instability has saturated into the
rotating plasma channel. This experiment could be proposed as a 
test-experiment for the analysis 
of the stability of collisional magnetoplasmas created by electron 
beams, in particular when turbulent states are studied. In that case, 
the understanding of the equilibrium of the magnetoplasma 
has to take into account the intermittent expulsion of negatively 
charged plasma filaments.\\
\\
The authors are indebted to N. Claire, C. Rebont and A. Escarguel for stimulating 
discussions. The authors would like also to express their gratitude to 
Pr. M. Shoucri and Pr. F. Doveil for the critical reading of the manuscript and for 
useful suggestions. This work was supported by Centre National de la
Recherche Scientifique, R\'egion PACA and Conseil R\'egional des
Bouches du Rh\^one. The authors thank also the expert technical
assistance from G. Vinconneau, A. Ajendouz and K. Quotb.
\vfill

\end{document}